# Complexity Analysis of 2-Heterogeneous Minimum Spanning Forest Problem

Zhujun Zhang

Qiang Sun

East China Normal University
Department of Computer Science and Technology
Shanghai, China
e-mail: zhangzhujun1988@163.com

*Abstract*—For complexity of the heterogeneous minimum spanning forest problem has not been determined, we reduce 3-SAT which is NP-complete to 2-heterogeneous minimum spanning forest problem to prove this problem is NP-hard and spread result to general problem, which determines complexity of this problem. It provides a theoretical basis for the future designing of approximation algorithms for the problem.

## I. INTRODUCTION

Heterogeneous minimum spanning forest problem(HMSF) was introduced by Yadlapalli et al. [1], and approximation algorithm was designed for heterogeneous vehicle routing problem in [1]. Heterogeneity of an undirected weighted complete graph refers to that each edge in the graph possesses a number of different costs. The goal of HMSF is to search the minimum cost spanning forest in a heterogeneous graph. An approximate algorithm of HMSF was proposed in [1], but complexity of this problem is not clear[1][2]. The main contribution of this paper is to prove HMSF is NP-hard by reducing a well known NP-complete problem to HMSF in which each edge in graph possesses two costs, which determines the complexity of the problem.

## II. NOTATION

This section describes the notation used in the whole paper.

Undirected complete graph $G=(V,E)$ is heterogeneous if each edge in the graph possesses more than one non-negative integer costs. If each edge e in graph $G$ possesses exactly two non-negative integer costs then $G$ is 2-heterogeneous. Let $w_1(e)$ and $w_2(e)$ denote the costs, $w_1$ and $w_2$ are cost functions on edge set $E$. For any node $v_1, v_2, v_3$ in $V$, if cost function $w$ meets $w(v_1,v_2)+w(v_2,v_3) \geqslant w(v_1,v_3)$ then $w$ satisfies the triangle inequality.

Spanning forest $F$ in the graph $G$ consists of two disjoint trees $T_1$ and $T_2$, where $T_1$ and $T_2$ contain all nodes in the graph. The cost of edge in $T_1(T_2)$ is defined by function $w_1(w_2)$. The cost of tree $T_1(T_2)$ is the sum of costs of edges in tree $T_1(T_2)$. The cost of spanning forest $F$ is the sum of the costs of $T_1$ and $T_2$.

2-Heterogeneous minimum spanning forest problem(2-HMSF) refers to search a minimum cost spanning forest in a 2-heterogeneous graph with given two nodes as tree roots. Determination form of 2-HMSF refers to that given a 2-heterogeneous graph, two nodes $r_1, r_2$ and a integer $k$, determinate whether there exists a spanning forest $F$ such that nodes $r_1$ and $r_2$ are roots of tree $T_1$ and $T_2$ respectively and the cost of $F$ is no larger than $k$.

3-SAT is a classical NP-complete problem, and it will be used in section 3. A formula is in 3-conjunctive normal form (3-CNF) if it is a conjunction of clauses, where a clause is a disjunction of three literals. For example, $(x_1 \vee \neg x_2 \vee x_3) \wedge (x_3 \vee \neg x_4 \vee \neg x_5)$ is in 3-CNF which contains two clauses and uses five variables. 3-SAT refers to determine whether a given formula in 3-CNF could be satisfied.

## III. PROOF

We first use reduction technique to prove 2-HMSF in general graph is NP-hard, and then explains how to use the same method in complete graph which satisfies triangle inequality. 3-SAT will be used as reduction problem. For any instance of 3-SAT, we construct a heterogeneous graph $G$, and specify the two nodes $r_1, r_2$ and integer $k$, then prove that the instance could be satisfied if and only if there exists a spanning forest $F$ in graph $G$ such that cost of $F$ is no larger than $k$ and nodes $r_1, r_2$ are tree roots.

Assume that the instance of 3-SAT contains m clauses $C_1, C_2, ..., C_m$ and uses n variables $x_1, x_2, ..., x_n$. Construct a 2-heterogeneous graph G as follows: For each variable $x_i$ in the instance of 3-SAT construct nodes $x_i$ and $\neg x_i$, and each clause $C_j$ construct a node $C_j$, then construct two nodes $t$ and $f$ represent true and false respectively; For each pair of nodes $x_i$ and $\neg x_i$, construct a edge $(x_i, \neg x_i)$, define the cost $w_1(x_i, \neg x_i)=w_2(x_i, \neg x_i)=1$, call these edges type x edges; Construct edge $(t, x_i)$ between node $t$ and each $x_i$, define the cost $w_1(t,x_i)=n+1$, $w_2(t,x_i)=(n+1)^2$, call these edges type t edges; Construct edge $(f, \neg x_i)$ between node $f$ and each $\neg x_i$, define the cost $w_1(f,\neg x_i)=(n+1)^2$, $w_2(f,\neg x_i)=n+1$, call these edges type f edges; For each clause $C_j$, construct three edges between the clause node and nodes corresponding three literals in $C_j$, for the edges of form $(C_j, x_i)$ define the cost $w_1(C_j, x_i)=(n+1)^2$, $w_2(C_j, x_i)=2(n+1)^2$, for the edges of form $(C_j, \neg x_i)$ define the cost $w_1(C_j, \neg x_i)=2(n+1)^2$, $w_2(C_j, \neg x_i)=(n+1)^2$, call these edges type C edges; Let node $t$ be root of tree $T_1$ and node $f$ be root of tree $T_2$, then let $k=m(n+1)^2+n(n+1)+n$.

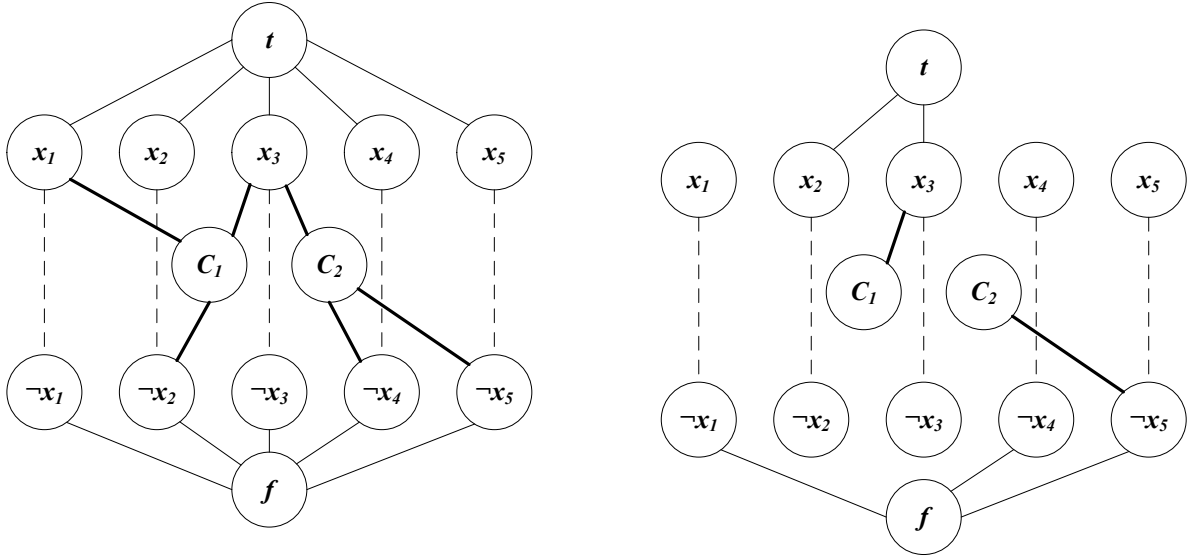

Figure 1. Example of 2-heterogeneous graph: (a) 2-heterogeneous graph (b) a spanning forest in 2-heterogeneous graph

Fig. 1(a) shows the 2-heterogeneous graph corresponding the formula $(x_1 \vee \neg x_2 \vee x_3) \wedge (x_3 \vee \neg x_4 \vee \neg x_5)$, where the dashed lines represent type x edges, and the thin solid lines represent type f edges, and the thick solid lines represent type C edges.

**Lemma 1**: Instance of 3-SAT could be satisfied if and only if there exists a spanning forest $F$ in graph $G$ constructed as above such that cost of $F$ is no larger than $k$ and nodes $t$ and $f$ are tree roots.

**Proof**: (necessity) Assume $\tau$ is a satisfying assignment of the formula. Construct a spanning forest $F$ with cost no larger than $k$, and at first $F$ does not contain any edges. For each variable $x_i$ in formula, if $\tau(x_i)=true$ than add edge $(t,x_i)$ to $F$, otherwise add edge $(f,\neg x_i)$ to $F$; Since the formula is satisfied under the assignment $\tau$, for each clause $C_j$, one could choose a literal in $C_j$ where true value of the literal is *true*, if literal $x_i$ is chosen than add edge $(C_j,x_i)$ to $F$, and if literal $\neg x_i$ is chosen than add edge $(C_j,\neg x_i)$ to $F$; Finally add all type x edges to $F$. It is easy to verify that $F$ is a spanning forest. Total cost of type C edges in $F$ is $m(n+1)^2$. Total cost costs of type t and type f edges in $F$ is $n(n+1)$. Total cost costs of type x edges in $F$ is $n$. Therefore the cost of forest $F$ is exactly $k$. Fig. 1(b) shows a spanning forest in a 2-heterogeneous graph.

(sufficiency) Assume $F$ is a spanning forest such that nodes $t$ and $f$ are tree roots and cost of $F$ is no larger than $k$. Construct a assignment $\tau$ as follows, for each variable $x_i$ in formula, if $F$ contains edge $(t,x_i)$ then define $\tau(x_i)=true$, otherwise define $\tau(x_i)=false$. We assert that assignment $\tau$ is valid and formula is satisfied under the assignment $\tau$.

We first prove formula is satisfied under assignment $\tau$. Consider type C edges, since $F$ is a spanning forest, $F$ must contain one of type C edges for each node $C_j$, therefore $F$ contains at least m type C edges. Since the cost of type C edges is either $(n+1)^2$ or $2(n+1)^2$, if $F$ contains more than m type C edges or edges with cost $2(n+1)^2$ then the cost of $F$ will be larger than $(m+1)(n+1)^2$, which is contrary to the assumption that cost of $F$ is no larger than $k$, therefore $F$ contains exactly m type C edges and the cost of each these edges is $(n+1)^2$. This means that each node $C_j$ corresponds to one of type C edges with cost $(n+1)^2$. Assume edge $(C_j,x_i)$ joins node $C_j$, then nodes $C_j$ and $x_i$ must be in tree $T_1$, otherwise cost of edge $(C_j,x_i)$ will be $2(n+1)^2$. Tree $T_1$ contains node $x_i$ implies that $F$ contains edge $(t,x_i)$, consequently clause $C_j$ is satisfied under assignment $\tau$ according to the definition of $\tau$ and the construction of graph $G$. Same consequence could be obtained when edge $(C_j,\neg x_i)$ joins node $C_j$. Therefore, each clause in the formula is satisfied under the assignment $\tau$, which implies the formula is satisfied under the assignment $\tau$.

Now we prove assignment $\tau$ is valid. Consider type t and type f edges in $F$, since $F$ is a spanning forest, for each pair of nodes $x_i$ and $\neg x_i$, $F$ contains at least one of edges $(t,x_i)$ and $(f,\neg x_i)$, otherwise nodes $x_i$ and $\neg x_i$ will not be in tree $T_1$ or $T_2$, thus $F$ contains at least n type t and type f edges. Discussion in last paragraph argues that total cost of type C edges in $F$ is $m(n+1)^2$, while cost of each type t and type f edges is no smaller than $(n+1)$. If $F$ contains more than n type t and type f edges, the cost of $F$ will be larger than $(m+1)(n+1)^2$, which is contrary to the assumption that cost of $F$ is no larger than $k$. Therefore $F$ contains exactly n type t and type f edges, and for each pair of nodes $x_i$ and $\neg x_i$, either edge $(t,x_i)$ or edge $(f,\neg x_i)$ is in $F$. Consequently assignment $\tau$ is valid from the definition of $\tau$. □

Obviously, the construction of the graph $G$ and reduction in Lemma 1 could be accomplished in polynomial time, and 3-SAT is NP-complete, thus 2-HMSF in general graph is NP-hard.

Consider 2-HMSF in complete graph. Add new edges in the heterogeneous graph $G$ constructed before to form new graph $G'$ such that $G'$ is a complete graph. In graph $G'$, costs of those edges in graph $G$ remain unchanged. For each newly added edge $e$ in $G'$, define $w_1(e)$ as the shortest distance of two vertices of e in the graph $G$ under cost function $w_1$, and the definition of $w_2(e)$ is similar. A spanning forest in graph $G$ clearly must be a spanning forest in graph $G'$. A spanning forest $F'$ in graph $G'$ could be converted into a spanning forest $F$ such that cost of $F$ is not larger than cost of $F'$, since those edges that are not in $G$ could be substituted by shortest path in $G$, and construction of $G'$ implies cost of $F$ is not larger than cost of $F'$. Therefore proof of Lemma 1 could also be used in complete graph, thus 2-HMSF in complete is NP-hard.

Consider 2-HMSF in complete graph which satisfies triangle inequality. In graph $G$ constructed before, redefine $w_2(C_j,x_i)$ and $w_1(C_j,\neg x_i)$ as $(n+1)^2+(n+1)$, and other costs remain unchanged, then expand graph $G$ to obtain complete graph $G'$ like in last paragraph. One can verify that cost functions $w_1$ and $w_2$ in graph $G'$ satisfy triangle inequality. Similarly, proof of Lemma 1 could also be used in graph $G'$, therefore 2-HMSF in complete graph which satisfies triangle inequality is NP-hard.

Discussion in last several paragraphs argues Theorem 1.

**Theorem 1**: 2-HMSF is NP-hard.

For determination form of 2-HMSF, there exists a simple polynomial time verification algorithm, so 2-HMSF∈NP. Combined with Theorem 1, 2-HMSF is NP-complete. For the general problem, k-HMSF(k⩾2) is also NP-complete obviously.

## IV. CONCLUSION

This paper presents a reduction from 3-SAT to 2-HMSF, which proves 2-HMSF is NP-hard. Actually determination form of 2-HMSF is NP-complete. Therefore there does not exist precise polynomial time algorithm for 2-HMSF, unless P=NP. Future research could focus on designing better approximation algorithm for 2-HMSF or analyzing approximability.

### REFERENCES

[1] S.K. Yadlapalli, S. Rathinam, S. Darbha, An approximation algorithm for a 2-Depot, heterogeneous vehicle routing problem, American Control Conference, 2009, pp. 1730-1735

[2] S. Yadlapalli, S. Rathinam, S. Darbha, 3-Approximation algorithm for a two depot, heterogeneous traveling salesman problem, Optimization Letters, vol. 6(1), 2012, pp. 141-152

[3] R. Bhattacharyya, A note on complexity of traveling tournament problem, Optimization Online, http://www.optimization-online.org/DB_FILE/2009/12/2480.pdf, 2009

[4] H. Yaman, Formulations and Valid Inequalities for the Heterogeneous Vehicle Routing Problem, Mathematical Programming, vol. 106(2), 2006, pp. 365-390

[5] S. Yadlapalli, J. Bae, S. Rathinam, S. Darbha, Approximation Algorithms for a Heterogeneous Multiple Depot Hamiltonian Path Problem, American Control Conference 2011, pp. 2789-2794

[6] S. O. Krumke, S. Saliba, T. Vredeveld, S. Westphal, Approximation algorithms for a vehicle routing problem, Mathematical Methods of Operations Research, vol. 68(2), 2008, pp. 333-359